\documentclass[aps,prb,reprint,superscriptaddress,showpacs]{revtex4-1}

\usepackage{graphicx}

\begin{document}

\title{Microscopic studies of the normal and superconducting state of Ca$_3$Ir$_4$Sn$_{13}$}

\author{S.~Gerber}
\email[]{simon.gerber@psi.ch}
\author{J.~L.~Gavilano}
\affiliation{Laboratory for Neutron Scattering, Paul Scherrer Institute, CH-5232 Villigen, Switzerland}
\author{M.~Medarde}
\affiliation{Laboratory for Developments and Methods, Paul Scherrer Institute, CH-5232 Villigen, Switzerland}
\author{V.~Pomjakushin}
\affiliation{Laboratory for Neutron Scattering, Paul Scherrer Institute, CH-5232 Villigen, Switzerland}
\author{C.~Baines}
\affiliation{Laboratory for Muon Spin Spectroscopy, Paul Scherrer Institute, CH-5232 Villigen, Switzerland}
\author{E.~Pomjakushina}
\author{K.~Conder}
\author{M.~Kenzelmann}
\affiliation{Laboratory for Developments and Methods, Paul Scherrer Institute, CH-5232 Villigen, Switzerland}

\date{\today}

\begin{abstract}
We report on muon spin rotation ($\mu$SR) studies of the superconducting and magnetic properties of the ternary intermetallic stannide Ca$_3$Ir$_4$Sn$_{13}$. This material has recently been the focus of intense research activity due to a proposed interplay of ferromagnetic spin fluctuations and superconductivity. In the temperature range $T=1.6-200$~K, we find that the zero-field muon relaxation rate is very low and does not provide evidence for spin fluctuations on the $\mu$SR time scale. The field-induced magnetization cannot be attributed to localized magnetic moments. In particular, our $\mu$SR data reveal that the anomaly observed in thermal and transport properties at $T^*\approx38$~K is not of magnetic origin. Results for the transverse-field muon relaxation rate at $T=0.02-12$~K, suggest that superconductivity emerges out of a normal state that is not of a Fermi-liquid type. This is unusual for an electronic system lacking partially filled $f$-electron shells. The superconducting state is dominated by a nodeless order parameter with a London penetration depth of $\lambda_{\rm L}=385(1)$~nm~and the electron-phonon pairing interaction is in the strong-coupling limit.
\end{abstract}


\maketitle

\section{Introduction} 

Interactions among charge, spin, orbital, and lattice degrees of freedom lead to emergent symmetry-breaking ground states in correlated electron systems, such as charge or spin order, superconductivity, and structural transitions.\cite{dagotto05} The correlations between the different degrees of freedom can lead to coexistence and even a coupling of the different constituents, such as superconductivity close to charge order\cite{morosan06} and magnetically mediated superconducting states.\cite{mathur98,monthoux07}

Ca$_3$Ir$_4$Sn$_{13}$ is a member of the material class of superconducting and/or magnetic ternary intermetallic stannides. The compound was first synthesized more than 30~years ago,\cite{remeika80,espinosa80,cooper80} but the condensed matter community has recently regained interest in it.\cite{yang10,hayamizu11,goh11,klintberg12,wang12,zhou12} Only a few of the physical properties, such as the superconducting transition at $T_{\rm c}\approx7$~K and the quasiskutterudite crystal structure, were reported in early studies.\cite{espinosa80,cooper80}

Recently an anomaly at $T^*\approx38$~K, well above the superconducting transition, was detected by macroscopic probes,\cite{yang10,wang12} such as magnetization $M(T)$, electrical resistivity $\rho(T)$, Seebeck coefficient, and Hall resistivity. Yang~\textit{et~al.}\cite{yang10} have proposed that the anomaly is the result of ferromagnetic spin fluctuations, coupled to superconductivity. Non-Fermi-liquid behavior was claimed from $\rho(T,B)$ at low applied magnetic fields with a crossover to a Fermi-liquid ground state as superconductivity is suppressed at high fields.\cite{yang10} Results for the electronic specific heat capacity $C_{\rm p}(T)$ suggest a nodeless\cite{wang12} strong-coupling\cite{hayamizu11} superconducting order parameter.

In the temperature-pressure phase diagram,\cite{goh11} a dome-shaped superconducting phase was found, along with a monotonous decrease in $T^*$ with increasing pressure $p$. A scenario where $T_{\rm c}$ and $T^*$ are linked by a $p$-induced superlattice quantum critical point, is supported by \mbox{x-ray} diffraction\cite{klintberg12} on the isovalent compound Sr$_3$Ir$_4$Sn$_{13}$. A structural transition is observed at $T^*$, involving a doubling of the unit cell ($Pm\overline{3}n$~$\rightarrow I\overline{4}3d$). The case of (Sr,Ca)$_3$Ir$_4$Sn$_{13}$ is reminiscent of that of Cu$_x$TiSe$_2$, where superconductivity emerges\cite{morosan06} as charge-density-wave (CDW) order is suppressed with increasing chemical doping $x$.

We have carried out muon spin rotation ($\mu$SR) experiments of Ca$_3$Ir$_4$Sn$_{13}$ to obtain additional insight into the role of ferromagnetic spin fluctuations in this material, especially in connection with the anomaly at $T^*\approx38$~K. The zero-field (ZF) muon relaxation rate is marginal in the temperature range $T=1.6-200$~K. We do not find an increase in the rate near $T^*$ that can be attributed to localized magnetic moments on the $\mu$SR time scale. This microscopic finding supports structural and/or CDW scenarios as the origin of the $T^*$~anomaly in ternary intermetallic stannides. We find no coupling of magnetic fluctuations and superconductivity. The nature of superconductivity, which emerges out of a normal state that is not of a Fermi-liquid type, was studied with low-temperature transverse-field (TF) measurements. Our results point to a nodeless strong-coupling superconductor with a London penetration depth of $\lambda_{\rm L}=385(1)$~nm.

\section{Sample characterization}

Gram-sized high-quality Ca$_3$Ir$_4$Sn$_{13}$ single crystals were synthesized using the Sn self-flux technique and concentrated HCl acid was used to remove excess Sn. A melting point of 1150$^\circ$~C was found using thermogravimetry. X-ray fluorescence reveals a homogeneous composition without segregations on a 20-$\mu$m length scale. 

\begin{figure}
	\includegraphics[width=\linewidth]{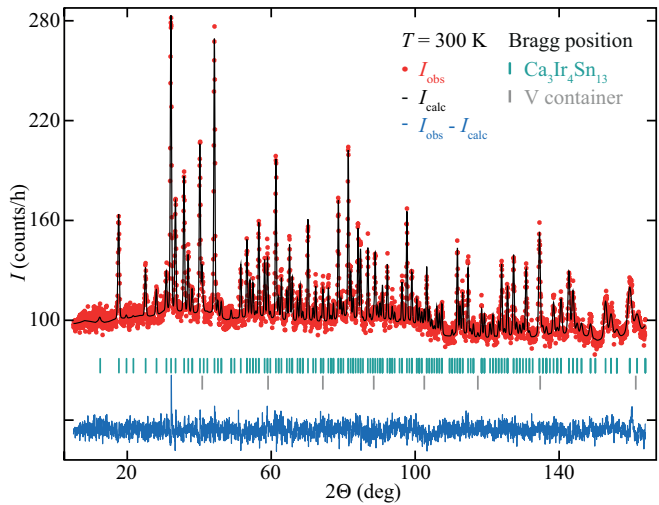}
	\caption{\label{hrpt}
		(Color online) Neutron powder diffraction of Ca$_3$Ir$_4$Sn$_{13}$ measured at $T=300$~K (red circles). The data is fitted with a single $Pm\overline{3}n$ phase and scattering from the vanadium sample container (black line). Green and grey dashes indicate Bragg positions of Ca$_3$Ir$_4$Sn$_{13}$ and vanadium, respectively. The largest differences between measured and refined intensities (blue line) appear from tiny shifts of the most intense Ca$_3$Ir$_4$Sn$_{13}$ Bragg peaks.}
\end{figure}

The crystal structure was investigated using neutron powder diffraction. Crushed Ca$_3$Ir$_4$Sn$_{13}$ single crystals were measured on the high-resolution powder diffractometer\cite{fischer00} (HRPT) at the Swiss Spallation Neutron Source SINQ, Paul Scherrer Institute, Villigen, Switzerland. The neutron diffraction pattern ($\lambda_{\rm n}=1.494$~\AA) measured at $T=300$~K indicates that the crystals are single phase (see Fig.~\ref{hrpt}). The data can be satisfactorily fitted using the cubic $Pm\overline{3}n$ structure, reported earlier,\cite{cooper80} yielding a lattice parameter of $a$~=~9.71922(8)~\AA.

\begin{figure}
	\includegraphics[width=\linewidth]{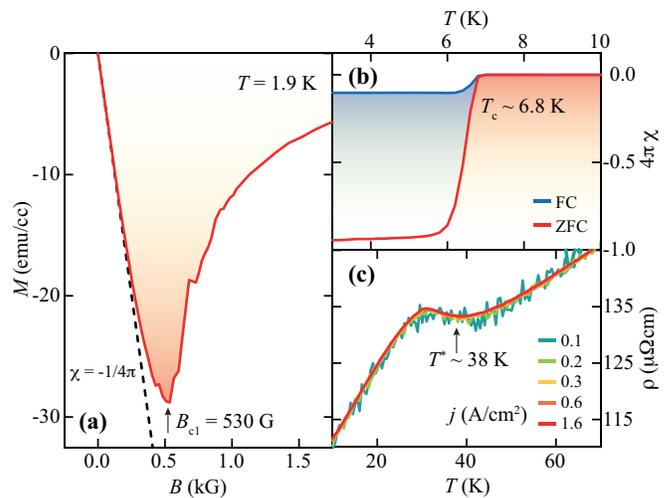}
	\caption{\label{ppms}
		(Color online) (a) Magnetization $M(B)$ as a function of the effective magnetic field $B$ at $T=1.9$~K. A diamagnetic Meissner state is found at low fields, with a crossover to the mixed state at the lower critical field $B_{\rm c1}=530$~G. (b) $T$~dependence of the ZFC and FC magnetic susceptibility $\chi(T)$ with the superconducting transition at $T_{\rm c}=6.8$~K. The difference between ZFC and FC points to vortex pinning. (c) An anomaly of the resistivity $\rho(T)$ is clearly visible at $T^*\approx38$~K for all applied current densities.}
\end{figure}

A physical properties measurements system (PPMS, Quantum Design) was used for an accurate characterization of macroscopic properties, such as $M(T,B)$ and $\rho(T)$. Figure~\ref{ppms}(a) shows the field dependence of the magnetization $M(B)$ at $T=1.9$~K, where $B$ is the effective field of the bulk sample, taking into account demagnetization effects. For low fields $B<200$~G we find a diamagnetic Meissner state ($\chi=$~-1/4$\pi$) with a crossover to the mixed state for $B$ greater than the lower critical field $B_{\rm c1}=530$~G. Superconductivity is type-II with the large value of $B_{\rm c2}/B_{\rm c1}\approx100$. 

Figure~\ref{ppms}(b) shows the magnetic susceptibility $\chi(T)$ (dc~measurement, $B_{\rm ext}=20$~G) with $T_{\rm c}\approx6.8$~K and a superconducting volume fraction of 94~\% at $T=3$~K after zero-field cooling (ZFC). Field cooling (FC) reduces $\chi(3$~K) to -$0.1/4\pi$, indicative of vortex pinning---most likely predominantly at the surface of the sample.

Our Ca$_3$Ir$_4$Sn$_{13}$ single crystals show a clear anomaly at $T^*\approx38$~K in $\rho(T)$ [see Fig.~\ref{ppms}(c)]. This anomaly was found in all of our samples and is thought to be intrinsic. Assuming a scenario where a pinned CDW emerges at $T^*$, one would expect nonlinear current-voltage $IV$ curves. This is the result of the expected depinning of the CDW for applied currents above a certain critical current density $j_{\rm c}$, which strongly depends on the actual pinning potential landscape. Thereby, the resistivity anomaly would be suppressed for sufficiently high applied currents, as observed\cite{monceau76} for NbSe$_3$. Our setup allowed the application of up to $j=1.6$~A/cm$^2$. However, at these current densities no change of the $T^*$~anomaly is observed.

\section{Experimental setup}

$\mu$SR experiments were carried out at the $\pi$M3~beamline of the Swiss Muon Source S$\mu$S, Paul Scherrer Institute, Villigen, Switzerland. Magnetic fields were applied along a cubic main crystal axis in all measurements.

High-temperature data ($T=1.6-200$~K) in ZF and a weak longitudinal field (wLF) were measured on the general-purpose-surface (GPS) muon spectrometer in the longitudinal muon polarization mode. A flow cryostat was used to control the temperature of one large Ca$_3$Ir$_4$Sn$_{13}$ single crystal ($m=~900$~mg) and a Helmholtz coil to apply $B_{\rm wLF}=30.5$~G. 

TF data were taken on the low-temperature facility (LTF) instrument in the transverse muon polarization mode. A dilution refrigerator and a superconducting magnet allowed access to the temperature range $T=0.02-12$~K with a longitudinal magnetic field of $B_{\rm TF}=622$~G. A mosaic of coaligned Ca$_3$Ir$_4$Sn$_{13}$ single crystals, glued onto an Ag sample plate, was measured on LTF.

\section{Results and discussion} 

\subsection{Search for ferromagnetic spin fluctuations}

\begin{figure}
	\includegraphics[width=\linewidth]{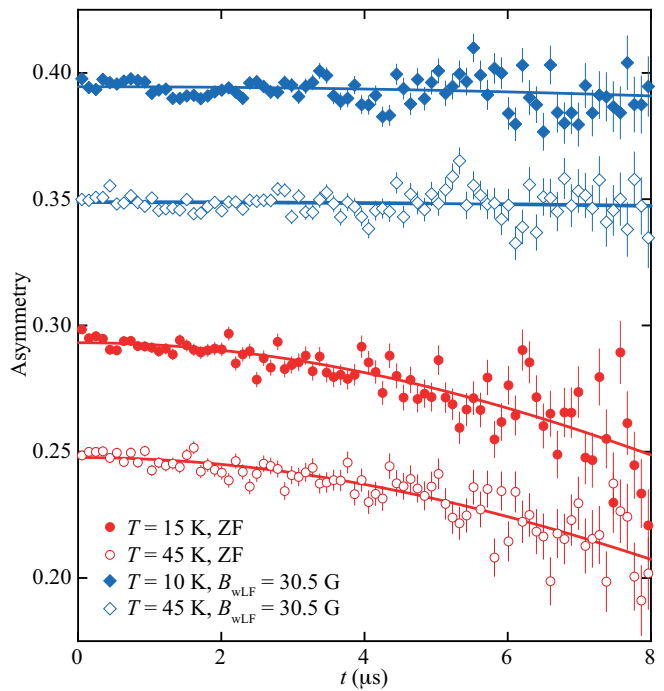}
	\caption{\label{gps_i}
		(Color online) Asymmetry $A(t)$ of ZF (circles) and wLF measurements (diamonds; $B_{\rm wLF}=30.5$~G) at temperatures below (filled symbols) and above (open symbols) $T^*\approx38$~K. The wLF quenches the nuclear moments, which reduces the muon relaxation rate. Lines depict fits to a static Gaussian Kubo-Toyabe relaxation function. The upper three data sets are shifted by 0.05 with respect to the lower series.}
\end{figure}

We carried out measurements of the ZF and wLF muon spin relaxation of longitudinally polarized muons to search for previously proposed ferromagnetic spin fluctuations.\cite{yang10} Figure~\ref{gps_i} shows the time dependence of the muon spin polarization $P(t)$ in the form of the asymmetry function
\begin{equation}
	A(t)=A_0\cdot P(t)=\frac{N_{\rm b}(t)-N_{\rm f}(t)}{N_{\rm b}(t)+N_{\rm f}(t)},
	\label{A}
\end{equation}
where $N_{\rm b}(t)$ and $N_{\rm f}(t)$ are the normalized histogram values of the backward and forward positron detector, respectively. ZF data are shown by circles, and wLF data with $B_{\rm wLF}=30.5$~G by diamonds. Open and filled data points were measured above and below $T^*$, respectively. The wLF quenches the randomly oriented, static (on muon time scales) nuclear dipole moments. The data in the time window $t=0-8~\mu$s were fitted in different manners and show only a marginal relaxation. In particular, the data are well described by a static Gaussian Kubo-Toyabe relaxation function,\cite{kubo81}
\begin{equation}
	P(t)=\frac{1}{3}+\frac{2}{3}\left(1-(\sigma t)^2\right)\exp\left(-\frac{(\sigma t)^2}{2}\right),
	\label{KT}
\end{equation}
where $\sigma$ is the muon relaxation rate due to quasistatic magnetic fields (see Fig.~\ref{gps_i}).

\begin{figure}
	\includegraphics[width=\linewidth]{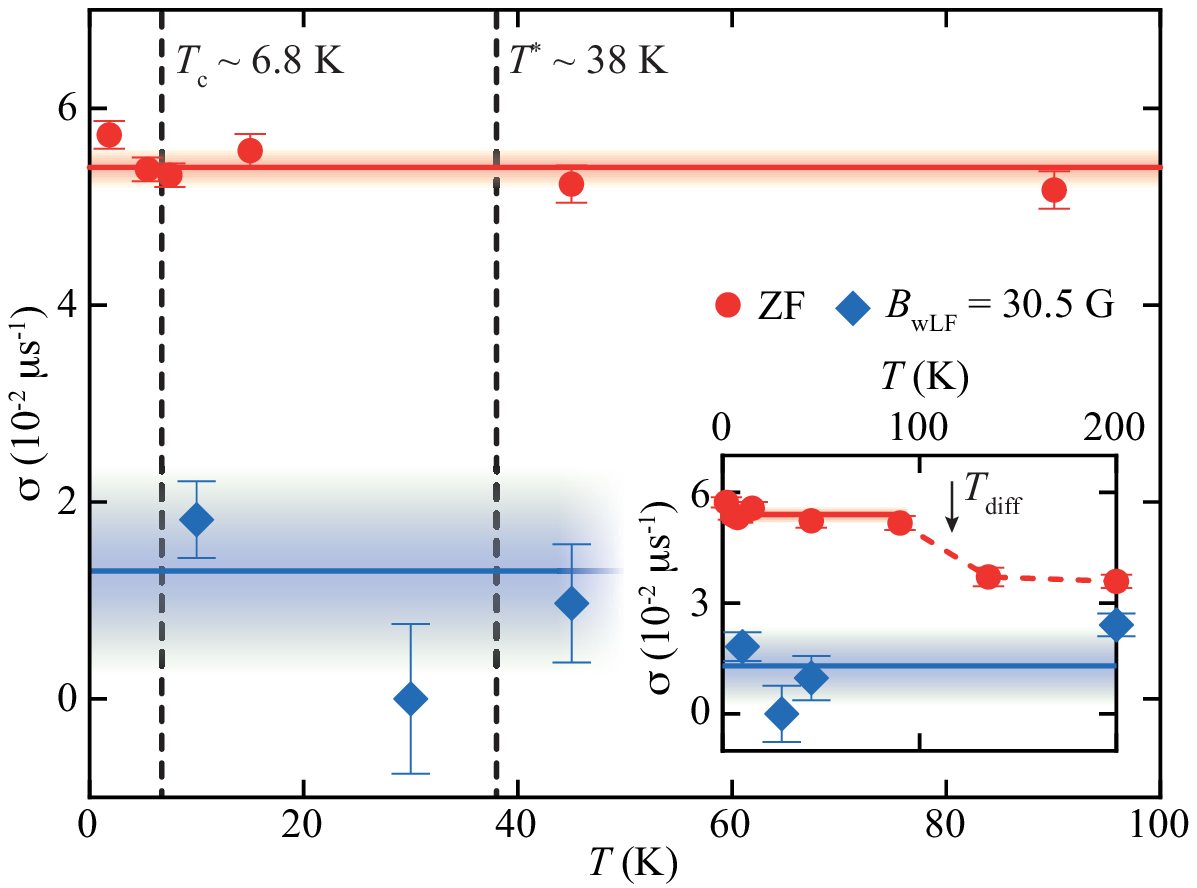}
	\caption{\label{gps_ii}
		(Color online) ZF (circles) and wLF (diamonds; $B_{\rm wLF}=30.5$~G) muon relaxation rate $\sigma(T)$ for $T=1.6-200$~K. We attribute the slight increase in $\sigma_{\rm ZF}$ at $T_{\rm diff}\approx120$~K to muon diffusion for $T>T_{\rm diff}$. In the range $T=1.6-100$~K, we find a constant low \mbox{$\sigma_{\rm ZF}=0.054(2)~\mu$s$^{-1}$} (red, shading: $\pm$~1~s.d.). This is supported by \mbox{$\sigma_{\rm wLF}=0.013(10)~\mu$s$^{-1}$} (blue), which is even lower and constant over the entire temperature range. An increase in the muon relaxation rate would be expected in a scenario with fluctuating or quasistatic magnetism at $T^*$.}
\end{figure}

Figure~\ref{gps_ii} shows the extracted muon relaxation rate $\sigma(T)$ of ZF and wLF measurements up to $T=100$~K. The full measured range $T=1.6-200$~K is shown in the inset. We find low $\sigma$ values of less than 0.06~$\mu$s$^{-1}$. $\sigma_{\rm ZF}$ is slightly enhanced for temperatures $T<T_{\rm diff}\approx120$~K. We attribute this effect to muon diffusion above $T_{\rm diff}$, leading to a decrease in the effective line width of the nuclear field distribution. Application of a weak longitudinal field $B_{\rm wLF}=30.5$~G suppresses, to a large extent, the already very low relaxation rate. $\sigma_{\rm wLF}=0.013(10)~\mu$s$^{-1}$ is constant over the entire temperature range $T=1.6-200$~K, and also $\sigma_{\rm ZF}=0.054(2)~\mu$s$^{-1}$ is constant for $T<100$~K. The error (shaded) areas in Fig.~\ref{gps_ii} correspond to 1 standard deviation (s.d.). From the ZF and wLF relaxation, we obtain an r.m.s. nuclear line width of $\sigma_{\rm nuc}\sim1$~G.

The muon relaxation rate does not change across the $T^*\approx38$~K anomaly, and also not at $T_{\rm c}$~=~6.8~K, as would be expected in a scenario involving fluctuating or quasistatic magnetism on the $\mu$SR time scale. We, therefore, conclude that our data do not support a scenario where spin fluctuations are present in Ca$_3$Ir$_4$Sn$_{13}$.

A Fermi surface reconstruction was conjectured from the $T$~dependence of the Seebeck coefficient.\cite{wang12} Our microscopic results put severe constraints on a magnetic model as the origin of the $T^*$~anomaly in the ternary intermetallic stannides, thereby, lending support to alternative nonmagnetic structural and/or CDW scenarios without the proposed coupling of magnetic fluctuations and superconductivity.

\subsection{Nature of the superconducting state}

\begin{figure}
	\includegraphics[width=\linewidth]{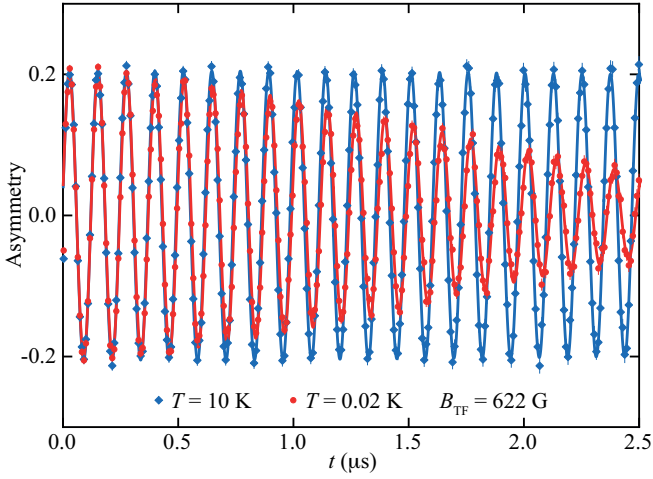}
	\caption{\label{ltf_i}
		(Color online) TF measurements with an FC applied magnetic field of $B_{\rm TF}=622$~G. Only a marginal relaxation is found in the normal state [$T=10$~K~$>T_{\rm c}$; (blue) diamonds]. However, $\sigma$ increases strongly deep inside the superconducting state [$T=0.02$~K~$\ll T_{\rm c}$; (red) circles] as a result of the field distribution from the ordered vortex lattice in the mixed state.}
\end{figure}

We carried out TF measurements with transverse muon polarization and an FC applied magnetic field of $B_{\rm TF}=622$~G~$>B_{\rm c1}$ to study the mixed/vortex state of superconducting Ca$_3$Ir$_4$Sn$_{13}$. The precession of the muon spin caused by $B_{\rm TF}$ is visible as an oscillation of $A(t)$ with a frequency $\gamma_{\mu}B_{\rm int}/2\pi$ (see Fig.~\ref{ltf_i}). $B_{\rm int}$ is the internal field of the bulk sample and $\gamma_{\mu}/2\pi=13.55$~kHz/G, with $\gamma_{\mu}$ the muon gyromagnetic ratio.\cite{sonier00} The left and right positron detectors were used for analysis of the LTF data to minimize absorption effects from the sample. The normal state data ($T=10$~K~$>T_{\rm c}$) show only a marginal relaxation, as expected from the ZF and wLF results. However, deep inside the superconducting state ($T=0.02$~K~$\ll T_{\rm c}$) the relaxation increases strongly. This is the result of the field distribution arising from the ordered vortex lattice in the mixed state.
The TF data can be well fitted for $T=0.02-12$~K with a sinusoidally oscillating Gaussian relaxation function,\cite{sonier00}
\begin{equation}
	P(t)=\exp\left(-\frac{(\sigma t)^2}{2}\right)\cos\left(\gamma_{\mu}B_{\rm int}t+\theta\right),
	\label{TF}
\end{equation}
where $\theta$ is the phase of the muon spin polarization with respect to the positron detectors at $t=0$. No background term for the Ag sample plate was needed for fitting, since the mosaic of coaligned single crystals covered practically the entire area of incoming muons and the sample thickness ensured that all muons were absorbed before reaching the sample plate.

\begin{figure}
	\includegraphics[width=\linewidth]{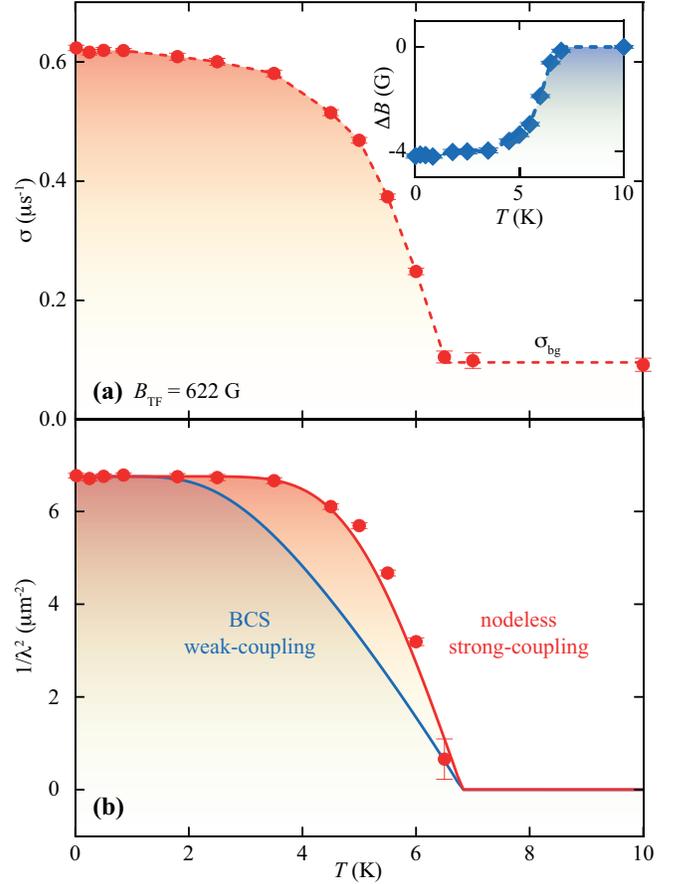}
	\caption{\label{ltf_ii}
		(Color online) (a) FC transverse-field muon relaxation rate $\sigma(T)=[\sigma^2_{\rm sc}(T)+\sigma^2_{\rm bg}]^{1/2}$, consisting of a superconducting contribution $\sigma_{\rm sc}(T)$ for $T<T_{\rm c}$ and a constant background $\sigma_{\rm bg}$. Inset: Diamagnetic shift $\Delta B(T)=B_{\rm int}(T)-B_{\rm eff}$. (b) $1/\lambda^2(T)\propto n_{\rm s}(T)$ with the weak-coupling prediction from BCS theory [lower (blue) curve] and a nodeless strong-coupling superconducting order parameter [upper (red) curve].}
\end{figure}

Figure~\ref{ltf_ii}(a) shows the TF muon relaxation rate \mbox{$\sigma(T)=[\sigma^2_{\rm sc}(T)+\sigma^2_{\rm bg}]^{1/2}$}, consisting of a constant nuclear background $\sigma_{\rm bg}=0.096~\mu$s$^{-1}$ and a superconducting component $\sigma_{\rm sc}(T)$ for $T<T_{\rm c}$. The $T$~dependence of the diamagnetic shift $\Delta B(T)=B_{\rm int}(T)-B_{\rm eff}$ with respect to the effective field $B_{\rm eff}$ of the bulk sample (averaged normal state $B_{\rm int}$) is shown in the inset in Fig.~\ref{ltf_ii}(a). Both $\Delta B(T)$ and $\sigma(T)$ are rather flat at very low temperatures, $T\ll T_{\rm c}$. This occurs when the superconducting gap function does not contain nodes, as is the case for $s$-wave pairing. The penetration depth $\lambda$ in the mixed state, assuming an ideal Ginzburg-Landau vortex lattice, can be derived\cite{brandt03} from $\sigma_{\rm sc}$ by
\begin{equation}
	\frac{1}{\lambda^2} = \frac{2\pi\sigma_{\rm sc}}{0.172\phi_0\gamma_{\mu}(1-b)[1+1.21(1-\sqrt{b})^3]},		\label{lambda}
\end{equation}
where $\phi_0=2.07\cdot10^{-11}$~G/m$^2$ is the magnetic flux quantum\cite{sonier00} and $b(T)=B_{\rm TF}/B_{\rm c2}(T)$. Experimental values of $B_{\rm c2}(T)$ were taken from Yang \textit{et al.} \cite{yang10}.

The penetration depth as derived from Eq.~(\ref{lambda}) is shown in Fig.~\ref{ltf_ii}(b) in the form of $1/\lambda(T)^2\propto n_{\rm s}(T)$, proportional to the superfluid density.\cite{sonier00} In the limit of zero temperature, we obtain a London penetration depth of \mbox{$\lambda_{\rm L}=\lambda(0)=385(1)$~nm}.
The lower (blue) curve shows the numerical solution of the weak-coupling BCS gap equation.\cite{bardeen57} In general, not just for weak-coupling, the $T$~dependence of the isotropic superconducting gap function can be approximated\cite{gross86} by
\begin{equation}
	\Delta(T)=\Delta_0\tanh\left(\frac{\pi}{\alpha}\sqrt{\frac{2}{3}\frac{\Delta C_{\rm p}}{\gamma T_{\rm c}}\left[\frac{T_{\rm c}}{T}-1\right]}\right),
	\label{Delta}
\end{equation}
with $\alpha=\Delta_0/k_{\rm B}T_{\rm c}$, $\Delta C_{\rm p}$ the jump of the specific heat capacity at $T_{\rm c}$, and $\gamma$ the Sommerfeld coefficient. The weak-coupling BCS superfluid density clearly does not account for the measured data. However, if the experimental value\cite{hayamizu11} $\Delta C_{\rm p}/\gamma T_{\rm c}=2.4$ is used in Eq.~(\ref{Delta}) and the coupling strength $\alpha$ is left as a free parameter, we find a lower bound of $\alpha\approx5$ [upper (red) curve in Fig.~\ref{ltf_ii}(b)]. From our microscopic data, we find an even stronger coupled nodeless superconducting order parameter than inferred from macroscopic measurements.\cite{hayamizu11} The derived $\alpha$ is larger than the corresponding values for Hg and Pb, typical strong-coupling electron-phonon superconductors.\cite{carbotte90}

The slight deviations between the experimental data and the approximation, Eq.~(\ref{Delta}), could in principle be accounted for by assuming a small admixture of a second superconducting gap. Such a scenario has been previously suggested.\cite{zhou12} However, our data rule out admixture of a large nodal gap, since low-energetic nodal excitations would lead to a nonzero slope of $n_{\rm s}(T)$ at the lowest temperatures---in contrast to our experimental results. Our data also rule out a high degree of admixture.

\subsection{$\mu$SR in the normal state}

\begin{figure}
	\includegraphics[width=\linewidth]{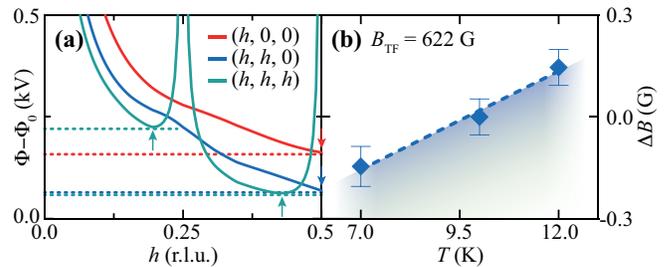}
	\caption{\label{NFL}
		(Color online) (a) Electrostatic potential along the $(h,~0,~0)$ (red), $(h,~h,~0)$ (blue), and $(h,~h,~h)$ (green) high-symmetry directions. Four candidates of muon sites are found at (1/2,~0,~0) (red arrow), (1/2~,1/2~,0) (blue arrow), and $(h,~h,~h)$ with $h=0.19(2)$ and 0.43(1) (green arrows). Ca$^{2+}$, Ir$^{4+}$, and Sn$^{4+}$ were assumed for the depicted calculation. (b)~A $T$-dependent diamagnetic shift in the normal state is not expected for a Fermi-liquid ground state.}
\end{figure}

We searched for the muon stopping sites by calculating the lattice sum of the electrostatic potential $\Phi$ at all ion positions (within a radius of 120~\AA) in the cubic unit cell. For all oxidation states, Ca$^{2+}$, Ir$^{2,3,4,6+}$, and Sn$^{2,4+}$, we find minima of $\Phi$ at the (1/2,~0,~0), (1/2~,1/2~,0) and $(h,~h,~h)$, with $h=0.19(2)$ and 0.43(1), positions and the crystallographically equivalent sites [see Fig.~\ref{NFL}(a)]. $\Phi_0$ in Fig.~\ref{NFL}(a) is a constant background to reduce the potential, as, for example, caused by the conduction electron's negative contribution.
For the four candidates of muon sites, we calculated the nuclear dipolar sum\cite{schenk85} to extract the expected r.m.s. $\mu$SR line width $\sigma_{\rm dip}$ (see Table~\ref{NFL_tab}). The analysis of the ZF and wLF data did not allow unambiguous determination of the muon stopping site.

\begin{table}
	\centering
	\caption{\label{NFL_tab}
		Minima of the electrostatic potential along high-symmetry $(h,~0,~0)$, $(h,~h,~0)$ and $(h,~h,~h)$ crystallographic directions with the calculated dipolar $\mu$SR line width $\sigma_{\rm dip}$.}
		\begin{tabular}{c | c | c}
ÊÊÊÊÊÊÊÊ		Direction	& $h$ (r.l.u.) & $\sigma_{\rm dip}$ (G)\\ \hline \hline
ÊÊÊÊÊ		$(h,~0,~0)$	& 1/2		& 1.2\\ \hline 
ÊÊÊÊÊ		$(h,~h,~0)$	& 1/2		& 0.9\\ \hline
ÊÊÊÊÊ		$(h,~h,~h)$	& 0.19(2)	& 0.9\\ 
ÊÊÊÊÊ					& 0.43(1)	& 2.1\\ 
	\end{tabular}
\end{table}

Figure~\ref{NFL}(b) shows the diamagnetic shift $\Delta B(T)$ in the normal state (TF measurements, $T=7-12$~K). Within the Fermi-liquid theory of metals\cite{ashcroft76} a constant $\Delta B$ is expected, as $\mu$SR locally probes contributions from (i) the paramagnetic Pauli susceptibility $\chi_{\rm P}$ and (ii) a diamagnetic van Vleck term from filled electronic shells. Both terms are expected to be $T$-independent. However, we observe a clear $T$~dependence of $\Delta B(T)$ at low temperatures but above $T_{\rm c}$. We attribute this to changes in $\chi_{\rm P}(T)$, which in turn reflects $T$-induced changes in the electronic density of states of the conduction electrons. This is not expected for a Fermi-liquid ground state. Strong electron-phonon coupling may affect the quasiparticles of the normal state, invalidating the Fermi-liquid concept in certain regimes.\cite{scalapino66} A scenario for such a strong electron-phonon coupling may, for example, arise from phonon softening phenomena.

Finally, we return to the question of localized magnetic moments. From the experimental $\mu$SR line width \mbox{$\sigma_{\rm nuc}\sim1$~G} (extracted from the ZF/wLF and the $T>T_{\rm c}$ TF data), we find no evidence of an appreciable contribution from possible localized electronic moments. In particular, we consider ferromagnetically ordered electronic moments, such as those deduced by Yang~\textit{et~al}.,\cite{yang10} resulting in internal fields \mbox{$B_{\mu}>1.4$~kG} at the candidates of muon sites, which is incompatible with the data. The local moments would have to be orders of magnitude smaller, $\mu_{\rm z}<0.001~\mu_{\rm B}$, to roughly correspond to the experimentally observed line width. Furthermore, our data do not reveal evidence of a relaxation due to fluctuating electronic moments on the $\mu$SR time scale. Therefore, we conclude that there are no localized magnetic moments at the Ir sites and we attribute the observed diamagnetic shift in the normal state to extended magnetic moments of the conduction electrons.

\section{Conclusions}

We have investigated the ternary intermetallic stannide Ca$_3$Ir$_4$Sn$_{13}$ using $\mu$SR. Over the entire range $T=1.6-200$~K we find no support for a scenario involving fluctuating or quasistatic localized magnetism on the $\mu$SR time scale. Therefore, our results favor a nonmagnetic scenario, where a structural transition and/or CDW formation occurs at $T^*$ in ternary intermetallic stannides. We have also studied the microscopic nature of the superconducting state of Ca$_3$Ir$_4$Sn$_{13}$, which emerges out of a normal state that is not of a Fermi-liquid type. We find a nodeless strong-coupling superconducting order parameter with a London penetration depth $\lambda_{\rm L}=385(1)$~nm. Our study may allow for admixture of a small second superconducting order parameter but rules out large nodal gaps.

\begin{acknowledgments}
Neutron powder diffraction and muon spin rotation experiments were performed at the Swiss Spallation Neutron Source SINQ and the Swiss Muon Source S$\mu$S, Paul Scherrer Institute, Villigen, Switzerland, respectively. We thank E.~Morenzoni for allocation of director's discretion time as well as A.~Amato and R.~Khasanov for support on GPS. Discussions with C.~Niedermayer and J.~S.~White are acknowledged. Neutron powder diffraction and $\mu$SR data were analyzed using the free software \textit{fullprof}\cite{rodriguez93} and \textit{musrfit}\cite{suter12}, respectively. This work was supported by the Swiss NSF (Contract No. 200021-122054, 200020-140345 and MaNEP).
\end{acknowledgments}

\end{document}